# Scalings for ultra-relativistic laser plasmas and monoenergetic electrons

S. Gordienko[1,2], A. Pukhov[1]

[1]*Institut für Theoretische Physik I, Heinrich-Heine-Universität Düsseldorf, D-40225, Germany*
[2]*L. D. Landau Institute for Theoretical Physics, Moscow, Russia*

(Dated: October 28, 2004)

The similarity theory is derived for ultra-relativistic laser-plasma interactions. First, it is shown that the most fundamental $S-$similarity is valid for both under- and overdense plasmas. Then, the particular case of tenious plasma is considered in great detail. It is shown that the electron dynamics in this case has two characteristic scales. The fast scale corresponds to relaxation to some attractor solution. The slow dynamics describes an adiabatic evolution of this attractor. This leads to a remarkable wave breaking exclusion rule in the $3D$ geometry. A similarity theory for the slow dynamics allows obtaining simple "engineering" scalings for the maximum electron energies, the number of accelerated electrons, the electron beam density, and for the acceleration distance. These scalings are aimed at design of a high-energy laser-plasma accelerator generating electron beams with superior properties.

## I. INTRODUCTION

The concept of laser-plasma electron acceleration has the decisive advantage over conventional accelerators: plasma supports electric fields orders of magnitude higher than the breakdown-limited field in radio-frequency cavities of conventional linacs. It is expected that the relativistic laser-plasma will finally lead to a compact high energy accelerator [1]. The very first experiments already have delivered high quality electron beams in the energy range 70...170 MeV [2–4]. Yet the way to a real laser-plasma accelerator that generates a high-energy electron beam with superior properties is long and full of problems which have to be solved. The main obstacle is that the experiments depend on too many parameters. Often, this makes the interpretation of experimental results ambiguous. At the same time, theoretical models suffer from the similar drawback. The system of kinetic equations describing the problem is both strongly non-linear and contains many parameters. As a result, the quest of searching for new perspective acceleration regimes is challenging and the physics of electron acceleration in plasma is often rather obscure.

The scientific difficulties just listed are neither new nor unique. A century ago hydrodynamics encountered and successfully overcame analogous difficulties. This has led not only to understanding of non-trivial theoretical problems, but also to a prominent practical achievement that is the design of the airplane. One can wonder how it was possible to design a reliable aircraft having neither complicated numerical simulations, nor comprehensive information about analytical solutions of the hydrodynamics equations. The answer is well known: it was the similarity theory that enabled physicists and engineers to find a clue to hydrodynamical problems [5, 6]. The similarity allows engineers to scale the behavior of a physical system from a laboratory acceptable size to the size of practical use.

To the best of our knowledge, no similarity theory has been applied to relativistic laser plasma interactions. This situation is surprising and unnatural, because the power of similarity theory for the magnetic confinement was recognized in the late 70s and the similarity theory [7] has been in use for design of large devices (tokamaks, stellarators) ever thereafter [8].

The aim of this article is to fill this theoretical gap. For the first time, we develop a similarity theory for laser-plasma interactions in the ultra-relativistic limit. Starting with the relativistic Vlasov equation on the electron distribution function and the Maxwell equations on electromagnetic fields, we show that the similarity parameter $S = n_e/a_0 n_c$ exists, where $a_0 = eA_0/m_e c^2$ is the relativistically normalized laser amplitude, $n_e$ is the plasma electron density and $n_c = m_e \omega_0^2/4\pi e^2$ is the critical density for a laser with the carrier frequency $\omega_0$. The basic ultra-relativistic similarity states that laser-plasma interactions with different $a_0$ and $n_e/n_c$ are similar as soon as the similarity parameter $S = n_e/a_0 n_c = const$ for these interactions. The $S-$similarity is so fundamental that it can serve as a crucial test to check reliability of various numerical codes and over-simplified theoretical models of relativistic laser-plasmas.

The basic $S-$similarity is valid for both over- and underdense plasmas. In the present work, we are interested in the special limit $S \ll 1$ of relativistically underdense plasmas as it is important for the high energy electron acceleration. We develop a detailed similarity theory for this case and derive scalings for laser-plasma interactions where the parameter $S$ varies. It follows from the theory that in the optimal configuration the laser pulse has the focal spot radius $k_p R \approx \sqrt{a_0}$ and the duration $\tau \leq R/c$. Here, $k_p = \omega_p/c$ is the plasma wavenumber and $\omega_p^2 = 4\pi n_e e^2/m_e$ is the plasma frequency. This corresponds to the so called "Bubble" acceleration regime [9]. It appears that the bubble is a natural attractor for the laser-plasma dynamics.

The theory developed in the present paper reveals two characteristic time scales. The fast time scale describes relaxation to the attractor solution. The slow time scale corresponds to an adiabatic evolution of the attractor. We were able to develop a similarity theory for the slow dynamics and to obtain scalings important for practical realization of a possible future high energy laser-plasma accelerator.

The central result of our theory is that the bubble regime of electron acceleration is stable, scalable and the scaling for the maximum energy $E_{mono}$ of the monoenergetic peak in the electron spectrum is

$$E_{mono} \approx 0.65 m_e c^2 \sqrt{\frac{\mathcal{P}}{\mathcal{P}_{rel}}} \frac{c\tau}{\lambda}. \qquad (1)$$

Here, $\mathcal{P}$ is the laser pulse power, $\mathcal{P}_{rel} = m_e^2 c^5/e^2 \approx 8.5$ GW is the natural relativistic power unit, and $\lambda = 2\pi c/\omega_0$ is the laser wavelength. The scaling (1) assumes that the laser pulse duration satisfies the condition $c\tau < R$.

The scaling for the number of accelerated electrons $N_{mono}$ in the monoenergetic peak is

$$N_{mono} \approx \frac{1.8}{k_0 r_e} \sqrt{\frac{\mathcal{P}}{\mathcal{P}_{rel}}}, \qquad (2)$$

where $r_e = e^2/m_e c^2$ is the classical electron radius, and $k_0 = 2\pi/\lambda$.

It follows from the scalings (1)-(2) that the laser energy conversion efficiency $\eta$ into the monoenergetic electrons

$$\eta = \frac{N_{mono} E_{mono}}{\mathcal{P}\tau} \approx 20\% \qquad (3)$$

is a universal constant.

The parametric dependencies in the scalings (1)-(2) follow from the analytical theory while the numerical prefactors have been obtained from direct 3D Particle-in-Cell simulations. We collect some additional "engineering" scalings in section X.

The paper is organized as following. In section II, we derive the basic $S-$similarity of ultra-relativistic laser-plasma. In the following sections, we concentrate on the particular case of tenious plasmas with $S \ll 1$. In section III we elaborate a Hamiltonian approach to the electron dynamics in new canonical variables and introduce the axial gauge for potentials. Section IV discusses multiple time scale dynamics. In section V we compare wave breaking in the $3D$ and $1D$ geometries. It appears that the slow adiabatic dynamics has a lower dimensionality. This leads to a wave breaking exclusion rule in the $3D$ geometry. In section VI we construct the relevant set of three dimensionless parameters defined by the initial laser pulse. In section VII we develop similarity theory for the slow (adiabatic) dynamics. This similarity theory is of a non-trivial "ladder" type as clarified in section VIII. To check our analytical scalings we use 3D Particle-in-Cell (PIC) simulations using the code Virtual Laser-Plasma Laboratory (VLPL) [10]. The numerical results are discussed in section IX. The concluding section X recapitulates the main results and collects some additional simple scalings.

## II. BASIC ULTRA-RELATIVISTIC SIMILARITY

We consider collisionless laser-plasma dynamics and neglect the ion motion. The electron distribution function $f(t, \mathbf{r}, \mathbf{p})$ is described by the relativistic Vlasov equation

$$(\partial_t + \mathbf{v}\partial_\mathbf{r} - e(\mathbf{E} + \mathbf{v} \times \mathbf{B}/c) \partial_\mathbf{p}) f(t, \mathbf{p}, \mathbf{r}) = 0, \qquad (4)$$

where the self-consistent fields $\mathbf{E}$ and $\mathbf{B}$ satisfy the Maxwell equations [11]

$$\nabla_\mathbf{r} \cdot \mathbf{E} = 4\pi e(n_e + \rho), \quad \nabla_\mathbf{r} \cdot \mathbf{B} = 0, \qquad (5)$$
$$c\nabla_\mathbf{r} \times \mathbf{B} = 4\pi \mathbf{j} + 4\pi \partial_t \mathbf{E}, \quad c\nabla_\mathbf{r} \times \mathbf{E} = -\partial_t \mathbf{B},$$

where $n_e$ is the background electron density, $\rho = -\int f \, d\mathbf{p}$, $\mathbf{j} = -e\int \mathbf{v} f \, d\mathbf{p}$, and $\mathbf{p} = m_e \gamma \mathbf{v}$; $\gamma = [1 - (v/c)^2]^{-1/2}$ is the relativistic factor.

We suppose that the laser pulse vector potential at the time $t = 0$ short before entering the plasma is

$$\mathbf{A}(t=0) = \mathbf{a}\left((y^2 + z^2)/R^2, x/c\tau\right)\cos(k_0 x), \qquad (6)$$

where $k_0 = \omega_0/c$ is the wavenumber. For the envelope representation form (6) to have a sense, the focal spot size $R$ and the pulse duration $\tau$ must be large enough: $k_0 R \gg 1$ and $\omega_0 \tau \gg 1$.

Generally, if one fixes the laser envelope $\mathbf{a}(\mathbf{r}_\perp, x)$, then the laser-plasma dynamics depends on four dimensionless parameters: the laser amplitude $a_0 = \max|e\mathbf{a}/mc^2|$, the focal spot radius $k_0 R$, the pulse duration $\omega_0 \tau$ and the plasma density ratio $n_e/n_c$, where $n_c = m_e \omega_0^2/4\pi e^2$ is the critical density.

Now we are going to show that in the ultra-relativistic limit when $a_0 \gg 1$, the number of independent dimensionless parameters reduces to three: $k_0 R$, $\omega_0 \tau$ and $S$, where the similarity parameter $S$ is

$$S = \frac{n_e}{a_0 n_c}. \qquad (7)$$

Let us introduce the new dimensionless variables

$$\hat{t} = S^{1/2}\omega_0 t, \quad \hat{\mathbf{r}} = S^{1/2} k_0 \mathbf{r}, \quad \hat{\mathbf{p}} = \mathbf{p}/m_e c a_0, \qquad (8)$$
$$\hat{\mathbf{A}} = \frac{e\mathbf{A}}{mc^2 a_0}, \quad \hat{\mathbf{E}} = \frac{eS^{-1/2}\mathbf{E}}{mc\omega_0 a_0}, \quad \hat{\mathbf{B}} = \frac{eS^{-1/2}\mathbf{B}}{mc\omega_0 a_0},$$

and the new distribution function $\hat{f}$ defined as

$$f = \frac{n_e}{(m_e c a_0)^3} \hat{f}\left(\hat{t}, \hat{\mathbf{p}}, \hat{\mathbf{r}}, a_0, S, \hat{R}, \hat{\tau}\right), \qquad (9)$$

where

$$\hat{R} = S^{1/2} k_0 R, \quad \hat{\tau} = S^{1/2} \omega_0 \tau. \qquad (10)$$

The normalized distribution function $\hat{f}$ is a universal one describing the interaction of the given laser pulse with a steep plasma profile. The function $\hat{f}$ satisfies equations

$$\left[\partial_{\hat{t}} + \hat{\mathbf{v}} \partial_{\hat{\mathbf{r}}} - \left(\hat{\mathbf{E}} + (\hat{\mathbf{v}} \times \hat{\mathbf{B}})\right) \partial_{\hat{\mathbf{p}}}\right] \hat{f} = 0, \qquad (11)$$

$$\nabla_{\hat{\mathbf{r}}} \cdot \hat{\mathbf{E}} = 4\pi(1 + \hat{\rho}), \quad \nabla_{\hat{\mathbf{r}}} \cdot \hat{\mathbf{B}} = 0, \qquad (12)$$
$$\nabla_{\hat{\mathbf{r}}} \times \hat{\mathbf{B}} = 4\pi \hat{\mathbf{j}} + \partial_{\hat{t}} \hat{\mathbf{E}}, \quad \nabla_{\hat{\mathbf{r}}} \times \hat{\mathbf{E}} = -\partial_{\hat{t}} \hat{\mathbf{B}},$$

where $\hat{\mathbf{v}} = \hat{\mathbf{p}}/\sqrt{\hat{\mathbf{p}}^2 + a_0^{-2}}$, $\hat{\rho} = -\int \hat{f} d\hat{\mathbf{p}}$, $\hat{\mathbf{j}} = -\int \hat{\mathbf{v}} \hat{f} d\hat{\mathbf{p}}$ and the initial condition for the vector potential is

$$\hat{\mathbf{A}}(\hat{t} = 0) = \hat{\mathbf{a}}\left((\hat{y}^2 + \hat{z}^2)/\hat{R}, \hat{x}/\hat{\tau}\right) \cos\left(S^{-1/2} \hat{x}\right), \qquad (13)$$

with the slow envelope $\hat{\mathbf{a}}$ such that $\max|\hat{\mathbf{a}}| = 1$.

Eqs. (11) together with the initial condition (13) still depend on the four dimensionless parameters $\hat{R}$, $\hat{\tau}$, $S$ and $a_0$. However, the parameter $a_0$ appears only in the expression for the electron velocity $\hat{\mathbf{v}} = \hat{\mathbf{p}}/\sqrt{\hat{\mathbf{p}}^2 + a_0^{-2}}$. In the limit $a_0 \gg 1$ one can write

$$\hat{\mathbf{v}} = \hat{\mathbf{p}}/|\hat{\mathbf{p}}|.$$

Consequently, for the ultra-relativistic amplitude $a_0 \gg 1$, the laser-plasma dynamics does not depend separately on $a_0$ and $n_e/n_c$. Rather, they converge into the single similarity parameter $S$.

The ultra-relativistic similarity means that when the plasma density and the laser amplitude change simultaneously so that $S = n_e/a_0 n_c = const$, the laser-plasma dynamics remains similar. Particularly, this basic ultra-relativistic scaling states that for different interaction cases with the same $S = const$, the plasma electrons move along similar trajectories and their momenta $\mathbf{p}$ scale as

$$\mathbf{p} \propto a_0. \qquad (14)$$

Simultaneously, the number of electrons participating in this motion scales as

$$N_e \propto a_0. \qquad (15)$$

Self-consistent fields and potentials generated in the plasma scale as

$$\phi, \mathbf{A}, \mathbf{E}, \mathbf{B} \propto a_0 \qquad (16)$$

for $\omega_0 \tau = const$, $k_0 R = const$ and $S = const$.

The ultra-relativistic similarity is valid for arbitrary $S-$values. In fact, the $S$ parameter appears only in the initial condition (13). It plays the role of the laser frequency and separates the relativistically overdense plasmas with $S \gg 1$ from the underdense ones with $S \ll 1$. In the further discussion we concentrate on the special case of electron acceleration in underdense plasma, particularly the bubble regime [9]. A numerical check of the obtained scalings will be presented in section IX.

### III. AXIAL GAUGE AND THE CANONICAL TRASFORMATION

We begin to consider the special case of tenious plasma $S \ll 1$. The main application here is electron acceleration in laser-generated wake fields. Because $S$ is now a small parameter, one can hope to obtain additional scalings and to get a deeper insight into the physics of electron acceleration in underdense plasmas.

We begin our study with the Vlasov equation (11). Its characteristics are described by the Hamiltonian

$$\hat{H} = \sqrt{\left(\hat{\mathbf{p}} + \hat{\mathbf{A}}\right)^2 + a_0^{-2}} - \hat{\varphi},$$

where $\left(\hat{\varphi}, \hat{\mathbf{A}}\right)$ is the four-potential describing electric and magnetic fields $\hat{\mathbf{E}}$ and $\hat{\mathbf{B}}$.

To obtain useful analytical results, we use the axial gauge $\hat{\varphi} = \hat{A}_x$ and make a convenient canonical transformation [12]. We mention that the identity

$$\hat{p}_x d\hat{x} - \hat{H} d\hat{t} = \left(\hat{p}_x - \hat{H}\right) d\hat{x} - \hat{H} d\left(\hat{t} - \hat{x}\right)$$

is valid. Introducing the new time variable

$$T = \hat{t} - \hat{x} \qquad (17)$$

one concludes that the new momenta

$$P_0 = \hat{p}_x - \hat{H}, \quad P_1 = \hat{p}_y \quad P_2 = \hat{p}_z \qquad (18)$$

and the new coordinates

$$\mathbf{X} = (x_0, x_1, x_2) = (\hat{x}, \hat{y}, \hat{z}) \qquad (19)$$

are canonical variables corresponding to the new Hamiltonian

$$H(T, \mathbf{P}, \mathbf{X}) = \hat{H}(\hat{t}, \hat{\mathbf{p}}, \hat{\mathbf{r}})$$
$$H = -\frac{\boldsymbol{\pi}^2 + a_0^{-2}}{2P_0} - \frac{P_0}{2} - \hat{\varphi}. \qquad (20)$$

Here we introduce notations $\boldsymbol{\pi} = (P_1 + A_1, P_2 + A_2)$, and $(A_1, A_2) = \left(\hat{A}_y, \hat{A}_z\right)$.

The Hamiltonian (20) allows one to obtain evolution equations on $\mathbf{P}$ and $\mathbf{X}$. We write these dynamical equations in the form

$$\frac{dx_0}{dT} = Q_0, \qquad \frac{dx_i}{dT} = Q_i,$$
$$\frac{dP_0}{dT} = U_0, \qquad \frac{dQ_i}{dT} = \frac{U_i}{P_0}, \qquad (21)$$

where $i, j = 1, 2$, and

$$\begin{aligned}
Q_i &= (P_i + A_i)/P_0, \\
Q_0 &= 0.5\left(Q_1^2 + Q_2^2 + a_0^{-2}P_0^{-2} - 1\right), \\
U_0 &= Q_i \partial_{x_0} A_i + \partial_{x_0}\hat{\varphi} \\
U_j &= Q_i \partial_{x_j} A_i - Q_i \partial_{x_i} A_j + U_0 \partial_{x_0} A_j \\
&\quad + \partial_T A_j - Q_j \partial_{x_0}\hat{\varphi} + \partial_{x_j}\hat{\varphi} - Q_j Q_i \partial_{x_0} A_i.
\end{aligned} \qquad (22)$$

Using the new Hamiltonian (20) and the equations of motion (21) we find the new form of the Vlasov equation

$$\left(\partial_T + \hat{L} + \frac{1}{P_0}\partial_{Q_i}U_i\right)F = 0, \qquad (23)$$

where the operator $\hat{L} = Q_0 \partial_{x_0} + Q_i \partial_{x_i} + U_0 \partial_{P_0}$, and the distribution function $F = |J|\hat{f}$. The Jacobian $J$ of the transformation $(\mathbf{X}, P_0, Q_1, Q_2) \to (\hat{\mathbf{x}}, \hat{\mathbf{p}})$ is

$$|J| = \frac{2P_0^4 a_0^2}{P_0^2 a_0^2(\mathbf{Q}^2 + 1) + 1}.$$

We have deliberately used the non-canonical variables $(P_0, Q_1, Q_2, x_0, x_1, x_2, T)$ in the new form of the Valsov Eq. (23). The advantage of these variables is that they help to separate fast and slow electron dynamics in the laser pulse.

## IV. MULTIPLE TIME SCALES AND THE ATTRACTOR SOLUTION

In this section we show that the fast electron dynamics is in fact a relaxation to some attractor solution, while the slow dynamics describes an adiabatic evolution of this attractor.

In the ultra-relativistic limit we get $|P_i + A_i| \gg a_0^{-1}$. This means that the "physical" electron momentum is much larger than $m_e c$. In this limit, one can write

$$Q_0 = 0.5(Q_1^2 + Q_2^2). \qquad (24)$$

In addition, we mention that electrons at rest have $P_0 = -1/a_0$. This value of $P_0$ is small and can be considered as the initial condition for unperturbed plasma in front of the laser pulse. The inequality $|P_0| \ll 1$ still holds for not too large "times" $T$. We are interested exactly in this interaction stage and in the following analytics we use the relativistically simplified expression (24) for $Q_0$.

Using $|P_0| \ll 1$ as the small parameter, one can divide electron dynamics into the fast and the slow ones. Because of the factor $1/P_0$ in the right hand side of the last equation in (21), one concludes that this equation is responsible for the fast dynamics. In its turn, this equation leads to the last term in the Vlasov Eq. (23).

Let us now compare different terms in Eq. (23). We are interested only in the times $T$ for which $F$ is localized in the area $|P_0| \ll 1$. Integrating Eq. (23) over $P_0$ one easily obtains that the term $\int (1/P_0) \partial_{Q_i} U_i F \, dP_0$ responsible for the fast dynamics is much larger than the term $\int \hat{L} F \, dP_0$. Thus, we may neglect the term $\hat{L} F$ in Eq. (23) and consider the reduced Vlasov equation

$$\left(\partial_T + \frac{1}{P_0}\partial_{Q_i}U_i\right)F = 0, \qquad (25)$$

which describes the fast electron dynamics.

Eq. (25) has the form of a transport equation. To study its properties, we introduce the flow vector $\mathbf{V} = (V_1, V_2)$, where $V_1 = -U_1$, $V_2 = -U_2$, on the plane $\mathbf{Q} = (Q_1, Q_2)$. It is convenient to choose the "−" sign in the definition of $\mathbf{V}$ since $P_0 < 0$. It appears that the flow $\mathbf{V}$ contains two and only two critical points for any $X, x_i$: one stable and one unstable.

The critical points of the vector field $\mathbf{V}$ are defined as $\mathbf{V} = 0$, that is $V_1 = V_2 = 0$. A trivial analysis shows that the solutions of the equations $V_1 = 0$ and $V_2 = 0$ are hyperbolas whose asymptotes are perpendicular to each other (see Fig. 1). Moreover, the asymptotes of $V_2 = 0$ can be obtained from those of $V_1 = 0$ by a $\pi/4$ turn around the center of the $V_1 = 0$-hyperbola, Fig. 1. Thus, the vector field $\mathbf{V}$ has two critical points. We denote the stable (attracting) point $\mathcal{A}$ and the unstable (repulsing) one $\mathcal{R}$.

To investigate the type of the critical points $\mathcal{A}$ and $\mathcal{R}$ we linearize the vector field $\mathbf{V}$ in their vicinity:

$$V_i = \frac{\partial V_i(\mathcal{A}, \mathcal{R})}{\partial Q_j}\delta Q_j(\mathcal{A}, \mathcal{R}),$$

where $\delta Q_j(\mathcal{A}, \mathcal{R}) = Q_j - Q_j(\mathcal{A}, \mathcal{R})$.

¿From the symmetry of Fig. 1 it follows that

$$\frac{\partial V_i(\mathcal{A})}{\partial Q_j} = -\frac{\partial V_i(\mathcal{R})}{\partial Q_j},$$

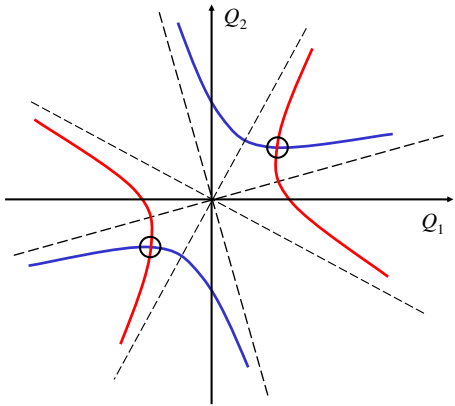

FIG. 1: Schematic representation of hyperbolas which are solutions of equations $V_1 = 0$ and $V_2 = 0$, where the vector **V** corresponds to the flow defined by Eq. (25). The intersections of these hyperbolas (marked by circles) correspond to the two critical points $\mathcal{A}$ and $\mathcal{R}$.

$i, j = 1, 2$. As a result, the characteristic numbers $\lambda_{1,2}(\mathcal{A}, \mathcal{R})$ of the critical points $\mathcal{A}$ and $\mathcal{R}$ satisfy the equalities

$$\lambda_{1,2}(\mathcal{A}) = -\lambda_{1,2}(\mathcal{R}).$$

Moreover, since

$$\alpha(\mathcal{A}, \mathcal{R}) = \partial V_1(\mathcal{A}, \mathcal{R})/\partial Q_1 = \partial V_2(\mathcal{A}, \mathcal{R})/\partial Q_2,$$
$$\beta(\mathcal{A}, \mathcal{R}) = -\partial V_1(\mathcal{A}, \mathcal{R})/\partial Q_2 = \partial V_2(\mathcal{A}, \mathcal{R})/\partial Q_1,$$

direct calculations give

$$\lambda_{1,2}(\mathcal{A}, \mathcal{R}) = \alpha(\mathcal{A}, \mathcal{R}) \pm i\beta(\mathcal{A}, \mathcal{R}), \quad (26)$$

where

$$\alpha = \partial_{x_0}\hat{\varphi} + \mathbf{Q}\partial_{x_0}\mathbf{A}, \quad (27)$$
$$\beta = \partial_{x_2}A_1 - \partial_{x_1}A_2 + Q_1\partial_{x_0}A_2 - Q_2\partial_{x_0}A_1. \quad (28)$$

In this formulas, $Q_i$, $A_i$ and all the derivatives are taken at the points $\mathcal{A}$ or $\mathcal{R}$ respectively.

Eqs. (27) state that the critical points are foci, and the relation (26) states that one of these foci is stable (attracting), while another is unstable (repulsing). The foci are counterrotating. We denote coordinates of the stable focus $\mathbf{Q}(\mathcal{A})$, and those of the unstable one $\mathbf{Q}(\mathcal{R})$.

The fast dynamics leads to a relaxation to the attracting focus. As soon as the attractor $\mathbf{Q}(\mathcal{A})$ is reached, one can consider the further slower dynamics as an adiabatic change of this attractor. Mathematically this means that the solution of Eq. (23) can be represented as

$$F = F_0(T, P_0, \mathbf{X})\delta(\mathbf{Q} - \mathbf{Q}(\mathcal{A})). \quad (29)$$

Here, the function $F_0$ does not depend on **Q**.

The fast and the slow times in our theory differ by the factor $|P_0| \propto 1/a_0 \ll 1$. In the further analysis we will see that the characteristic scales of the slow dynamics is $\Delta T_s \propto S^{1/2}$. This corresponds to the "physical" time $\Delta t_s \propto a_0 k_0/ck_p^2$. Thus, the characteristic scale of the fast time is $\Delta T_f \propto |P_0|S^{1/2}$. This corresponds to the "physical" time $\Delta t_f \propto |P_0|\Delta t_s \geq k_0/ck_p^2$. One can say that the attractor solution is reached after the laser pulse has propagated in plasma the characteristic distance $L_a \propto k_0/k_p^2$.

## V. 3D SELF-INJECTION AND WAVE BREAKING

In this section we demonstrate that the physics of ultra-relativistic $3D$ wave breaking differs qualitatively from that of $1D$ wave breaking. To emphasize this difference we consider the $3D-$adiabatic dynamics and compare the results with the $1D-$dynamics.

### A. 3D wave breaking exclusion rule

From Eqs. (23) and (29) one finds that the function $F_0$ satisfies the equation

$$\left(\partial_T + \hat{L}_{ad}\right) F_0 = 0, \quad (30)$$

where the operator

$$\hat{L}_{ad} = 0.5\mathbf{Q}^2(\mathcal{A})\partial_{x_0} + Q_i(\mathcal{A})\partial_{x_i} + U_0\partial_{P_0}. \quad (31)$$

describes the slow (adiabatic) dynamics. The characteristics of Eq. (31) are

$$\frac{dx_i}{dT} = Q_i(\mathcal{A}), \quad \frac{dx_0}{dT} = \frac{\mathbf{Q}^2(\mathcal{A})}{2}, \quad (32)$$
$$\frac{dP_0}{dT} = U_0, \quad (33)$$

Since $\mathbf{Q}(\mathcal{A})$ does not depend on $P_0$, one can solve the three Eqs. (32) first and then use this solution to obtain $P_0(T, \mathbf{X})$. Consequently, one writes for the function $F_0$

$$F_0 = n_e(T, \mathbf{X})\delta(P_0 - P_0(T, \mathbf{X})). \quad (34)$$

Here $P_0$ is a variable and $P_0(T, \mathbf{X})$ is obtained by integration of Eqs. (32).

According to Eqs. (29) and (34) the electron momentum at the point **X** and the "time" $T$ is unambiguously defined provided that

$$|P_0(T, \mathbf{X})| \ll 1. \quad (35)$$

The condition (35) means a remarkable "3D wave breaking exclusion rule" claiming that no wave breaking occurs in a 3D laser-plasma interaction as long as the inequality $|P_0(T, \mathbf{X})| \ll 1$ holds.

As the laser-plasma interaction goes on, the value of $|P_0|$ increases and finally the 3D-wavebreaking can occur. The condition for the possibility of the 3D-wavebreaking can be thus formulated as

$$|P_0| \geq 1. \tag{36}$$

The 3D-wavebreaking condition (36) should be compared with the standard 1D-wavebreaking condition.

### B. 1D wave breaking versus 3D-wave-breaking

The standard 1D wave breaking condition has been first formulated by Akhiezer and Polovin [14]. It claims that there is a limiting field $E_{WB}$ of a plasma wave running with the phase velocity $v_p$:

$$\frac{eE_{WB}}{mc\omega_p} = \sqrt{2(\gamma_p - 1)} \tag{37}$$

where $\gamma_p = 1/\sqrt{1 - (v_p/c)^2}$.

It is useful to present the 1D case in the same variables $(P_0, x_0, T)$ as defined in Eqs. (17)-(19). In the 1D geometry the Hamiltonian (20) takes on the very compact form

$$H = -\frac{1}{2a_0^2 P_0} - \frac{P_0}{2} - \hat{\varphi}. \tag{38}$$

For electron trajectories one obtains

$$\frac{dx_0}{dT} = v_0, \quad \frac{dP_0}{dT} = u_0, \tag{39}$$

where $v_0 = 1/(2a_0^2 P_0^2) - 1/2$ and $u_0 = \partial \hat{\varphi}/\partial x_0$.

Eqs. (39) yield the one-dimensional Vlasov equation in the form

$$\left( \partial_T + \partial_{x_0} v_0 + \partial_{P_0} u_0 \right) f = 0. \tag{40}$$

Now we are able to compare the 1D and 3D dynamics. Notice that for the 3D case we used the fact that the transverse momenta $|P_i + A_i| \gg a_0^{-1}$. That is why we neglected the term $a_0^{-2}/2P_0$ in the Hamiltonian (20). However this approximation cannot be done for the 1D-case (38), where this term plays the major role. Moreover, in the 3D-case we exploited the limit $|P_0| \ll 1$ to reduce the six equations (21) to the three closed equations (32). It is this dimensionality reduction that has lead to the 3D wave breaking exclusion rule (35).

In the 1D case one cannot reduce the two equations (39) even in the limit $|P_0| \ll 1$. As a result, the physical picture of the wavebreaking is very much different in the 1D and 3D cases. It is known that in the 1D case the wavebreaking leads to a multi-stream electron flow when one particle overtakes another. In the 3D-case the wave breaking looks rather like a thermalization because of an increase in the widths of $\delta(\mathbf{Q} - \mathbf{Q}(\mathcal{A}))$ and $\delta(P_0 - P_0(T, \mathbf{X}))$ in the electron distribution function. The characteristic "wavebreaking time" $t_b$ for thermalization is a time when $P_0$ becomes not small, $|P_0| \approx 1$.

Generaly speaking, there is no criterion relating the wavebreaking with the electric field value in a 3D plasma wave. In other words, the 3D geometry allows for regular (without wavebreaking) wake fields with arbitrary electric fields as long as the 3D wave breaking exclusion rule (35) holds.

## VI. PARAMETERS IN DYNAMIC EQUATIONS AND INITIAL CONDITIONS

In the previous section we have derived Eq. (30) that includes only the "slow" (adiabatic) dynamics. The "fast" dynamics putting the system into the vicinity of the attracting focus $\mathbf{Q}(\mathcal{A})$ has already been taken into account by Eq. (25). Now we are going to develop a similarity theory describing only this "slow" dynamics rather than the both "slow" and "fast" ones as Eq. (23) does. For this purpose, we introduce the following *tilde*-variables and *tilde*-functions

$$\begin{aligned}
\tilde{T} &= S^{-1/2}T, \quad \tilde{x}_0 = S^{1/2}x_0, \quad \tilde{x}_i = x_i, \\
\tilde{A}_i &= A_i, \quad \tilde{\varphi} = S^{1/2}\hat{\varphi}, \\
\tilde{Q}_i &= S^{1/2}Q_i, \quad \tilde{P}_0 = P_0, \quad \tilde{n}_e = S^{1/2}n_e,
\end{aligned} \tag{41}$$

where $i = 1, 2$.

We rewrite the Maxwell equations in the *tilde*-notation:

$$\begin{aligned}
(\partial^2_{\tilde{T}\tilde{x}_0} - S\partial^2_{\tilde{x}_0\tilde{x}_0} - \nabla^2_\perp)\tilde{\varphi} - \partial^2_{\tilde{T}\tilde{x}_j}\tilde{A}_j &= 4\pi(-S^{1/2} + \tilde{n}_e), \\
(\partial^2_{\tilde{T}\tilde{x}_0} - S\partial^2_{\tilde{x}_0\tilde{x}_0} - \nabla^2_\perp)\tilde{A}_i + \partial^2_{\tilde{x}_0\tilde{x}_i}\tilde{\varphi} + \partial^2_{\tilde{x}_i\tilde{x}_j}\tilde{A}_j &= \\
-8\pi \tilde{Q}_i \tilde{n}_e/(\tilde{Q}_1^2 + \tilde{Q}_2^2),
\end{aligned} \tag{42}$$

where $\nabla^2_\perp = \partial^2_{\tilde{x}_1\tilde{x}_1} + \partial^2_{\tilde{x}_2\tilde{x}_2}$.

Using Eqs. (42) we are now ready to exploit the limit $S \to 0$. In this limit we obtain

$$\begin{aligned}
(\partial^2_{\tilde{T}\tilde{x}_0} - \nabla^2_\perp)\tilde{\varphi} - \partial^2_{\tilde{T}\tilde{x}_j}\tilde{A}_j &= 4\pi \tilde{n}_e, \\
(\partial^2_{\tilde{T}\tilde{x}_0} - \nabla^2_\perp)\tilde{A}_i + \partial^2_{\tilde{x}_i\tilde{x}_j}\tilde{A}_j &= -8\pi \tilde{Q}_i \tilde{n}_e/(\tilde{Q}_1^2 + \tilde{Q}_2^2),
\end{aligned} \tag{43}$$

Differential operators in Eqs. (43) do not depend on any dimensionless parameter. Consequently, the retarded Green function of Eqs. (43) does not depend on any dimensionless parameter either. The same conclusion is valid for the retarded solution generated by the right hand side of Eqs. (43).

Now we have to take into account the initial condition (13) defined by the laser pulse. The four-potential of the electromagnetic field in plasma can be written as

$$(\varphi, \mathbf{A}) = (\varphi_h, \mathbf{A}_h) + (\varphi_r, \mathbf{A}_r), \qquad (44)$$

where $(\varphi_h, \mathbf{A}_h)$ is a homogeneous solution of the Maxwell equations, and $(\varphi_r, \mathbf{A}_r)$ is the retarded potential obtained by the convolution of the right hand side of Maxwell's equations with the retarded Green function [13].

Because all the dimensionless parameters have been shifted into the initial condition (13), they influence explicitely only the homogeneous solution $A_h$. We write it in the form

$$\mathbf{A}_h = \int \mathbf{a}_h(\omega, \mathbf{r}) \exp(-i\omega(t - x/c)) \, d\omega \, d\mathbf{r} + cc,$$

where $\mathbf{a}$ is the slow envelope the pulse.

Because $k_0 R \gg 1$, the $y$ and $z$ components of the slow envelope in the Lorentz gauge satisfy the Schrödinger equation

$$i\partial \mathbf{a}_h / \partial x = -(c/2\omega)(\partial^2/\partial y^2 + \partial^2/\partial z^2) \mathbf{a}_h.$$

Its solution is

$$\mathbf{A}_h^{y,z} = R^2 \tau \int \exp(i\sigma) \mathbf{u}(\tau(\omega - \omega_0), R\mathbf{k}_\perp, c\tau/R) \, d\omega \, d\mathbf{k}_\perp + cc,$$

where $\sigma = \omega(t - x/c) - c\mathbf{k}_\perp^2 x/(2\omega) + \mathbf{k}_\perp \mathbf{x}_\perp$; $\mathbf{k}_\perp = (k_y, k_z)$; $\mathbf{x}_\perp = (y, z)$. The function $\mathbf{u} = \mathbf{u}(\tau(\omega - \omega_0), R\mathbf{k}_\perp, c\tau/R)$ is generated by the slow envelope $\mathbf{a}$ of the laser pulse (6) so that

$$\mathbf{a} = R^2 \tau \int \exp(i\sigma) \mathbf{u}(\tau\omega, R\mathbf{k}_\perp) \, d\omega d\mathbf{k}_\perp.$$

The $x$-component of the vector potential can be found from the Lorentz condition $\nabla \cdot \mathbf{A}_h = 0$

$$A_h^x = -R^2 \tau \int \exp(i\sigma) \frac{(\mathbf{k}_\perp \cdot \mathbf{u})}{\omega} \, d\omega \, d\mathbf{k}_\perp + cc$$

and $\varphi_h = 0$.

The obtained Lorentz-potentials must be transformed to the axial gauge. The transformation is defined by the gauge function

$$g = 2iR^2 \tau \int \exp(i\sigma) \frac{c(\mathbf{k}_\perp \cdot \mathbf{u})}{\mathbf{k}_\perp^2} \, d\omega \, d\mathbf{k}_\perp + cc,$$

so that

$$\varphi \to -\frac{1}{c} \frac{\partial g}{\partial t}, \qquad \mathbf{A}_h \to \mathbf{A}_h + \nabla g.$$

Now we rewrite the homogeneous potentials in the *tilde*-notation. For transverse components of the vector potential one finds

$$\tilde{\mathbf{A}}_h^{y,z} = \int \exp(i\tilde{\sigma}) \tilde{\mathbf{u}} \, d\tilde{\omega} \, d\tilde{\mathbf{k}}_\perp + cc,$$

where

$$\begin{aligned}
\tilde{\sigma} &= \tilde{T} + \tilde{\omega} \tilde{T}/\Pi_3 - \tilde{\mathbf{k}}_\perp^2 \tilde{x}_0 / 2\Pi_1^2 + \tilde{\mathbf{k}}_\perp \tilde{\mathbf{x}}_\perp / \Pi_1, \quad (45) \\
\tilde{\mathbf{u}} &= \mathbf{u}(\tilde{\omega}, \tilde{\mathbf{k}}_\perp, \Pi_2)
\end{aligned}$$

This expression depends on the only three dimensionless parameters

$$\Pi_1 = \hat{R}, \quad \Pi_2 = \hat{\tau}/\hat{R}, \quad \Pi_3 = S^{-1/2} \hat{\tau},$$

The parameters $\Pi_1, \Pi_2, \Pi_3$ entirely define the laser-plasma dynamics.

The parameter $\Pi_1 = S^{1/2} k_0 R$ plays the role of the waist of the laser beam, the parameter $\Pi_2 = c\tau/R$ gives the laser pulse aspect ratio, and $\Pi_3 = \omega_0 \tau$ is the laser pulse duration.

## VII. SIMILARITY THEORY FOR THE SLOW DYNAMICS

One can easily realise that the homogeneous solution $\left(\hat{\varphi}_h, \hat{\mathbf{A}}_h\right)$ presented in the *tilde*-varialbes is the only source of dimensionless parameters on which the theory can depend (see, for example, Eq. (13)). This means that only if a dimensionless parameter enters $\left(\hat{\varphi}_h, \hat{\mathbf{A}}_h\right)$, this parameter is important for the theory. On the contrary, if the dependence of $\left(\hat{\varphi}_h, \hat{\mathbf{A}}_h\right)$ from a dimensionless parameter is negligible, the role of this parameter in the theory is negligible too. It is worth emphasizing that this fact is not related to the question where or how long the homogeneous solution exists in plasma. In other words, the homogeneous solution accumulates the entire informarion about initial conditions.

To develop a similarity theory we note that the parameter $\Pi_3$ gives the charateristic time of the slow envelope for the homogeneuos solution. If we are mainly interested in *tilde*-times

$$\tilde{T} < \Pi_3, \qquad (46)$$

then we can neglect the term with $\Pi_3$ in the homogeneous solution (45). Thus, our theory depends on the only two parameters $\Pi_1$ and $\Pi_2$. This allows for finding scalings describing the ultra-relativistic laser-plasma interaction.

First, we represent the characteristics equations (32)-(33) of the Vlasov Eq. (30) in the *tilde*-notation:

$$\frac{d\tilde{\mathbf{x}}}{d\tilde{T}} = \tilde{\mathbf{Q}}(\mathcal{A}), \quad \frac{d\tilde{x}_0}{d\tilde{T}} = \frac{\tilde{\mathbf{Q}}^2(\mathcal{A})}{2}, \qquad (47)$$

$$\frac{d\tilde{P}_0}{d\tilde{T}} = \frac{\Pi_1 \Pi_2}{\Pi_3} \tilde{U}_0, \qquad (48)$$

where $\tilde{U}_0 = -\tilde{Q}_i \partial_{\tilde{x}_0} \tilde{A}_i - \partial_{\tilde{x}_0} \tilde{\varphi}$, and $(\Pi_1 \Pi_2)/\Pi_3 = S^{1/2}$.

We see that Eqs. (47) do not contain the parameter $\Pi_3$ explicitely. On the other hand, the $P_0$ dynamics (48) is completely defined by the solution of Eqs. (47). This means that the solution for $P_0$ can be represented as

$$P_0 = S^{1/2} d_1(\Pi_1, \Pi_2, \tilde{T}, \tilde{\mathbf{x}}, \tilde{x}_0) \qquad (49)$$

where $d_1(\Pi_1, \Pi_2, \tilde{T}, \tilde{\mathbf{x}}, \tilde{x}_0)$ is a universal function that does not depend on $\Pi_3$.

### A. Scalings for the maximum electron energies

Now we are ready to obtain an expression for the relativistic $\gamma$-factor of electrons at the point $\tilde{x}_{0,1,2}$ at the "time" $\tilde{T}$. We use the Hamiltonian (20) and get

$$\gamma = -0.5 a_0 P_0 \mathbf{Q}^2(\mathcal{A}) = a_0 S^{-1/2} D_1(\Pi_1, \Pi_2, \tilde{T}, \tilde{\mathbf{x}}, \tilde{x}_0), \qquad (50)$$

where the functions $D_{1,2,3,...}$ (see also below) are universal.

The expression (50) describes the self-similar electron energy distribution. We are interested in the largest relativistic factor of electrons. Fixing the transverse coordinates at the axis $y = z = 0$, we obtain that after the accelerating distance $x$ the maximum $\gamma$-factor of electrons scales as

$$\gamma_m = a_0^{3/2} \sqrt{\frac{n_c}{n_e}} D_2\left(\Pi_1, \Pi_2, \frac{k_p^2 x}{a_0 k_0}\right), \qquad (51)$$

Using Eqs. (47)-(48) we come to the conclusion that

$$\frac{dP_0}{dx} = O\left(\frac{S^{1/2} k_p^2}{a_0 k_0}\right). \qquad (52)$$

Because of the 3D wavebreaking exclusion rule (35), the wave-breaking can take part only when $P_0$ becomes large, $|P_0| \approx 1$. According to (52) this happens at the distance

$$L_b = \frac{a_0 k_0}{S^{1/2} k_p^2} D_3(\Pi_1, \Pi_2), \qquad (53)$$

corresponding to the *tilde*-time

$$\tilde{T}_b = \Pi_3 / \Pi_2. \qquad (54)$$

Notice that in the case of "pancake"-like laser pulses with $\Pi_2 \leq 1$, the wavebreaking time (54) is larger than the maximum time (46) untill which our approximations are valid. Thus, we can describe the laser-plasma dynamics with $\Pi_2 < 1$ on the length

$$L_d = \Pi_2 L_b \qquad (55)$$

At the distance $L_d$, the variable $P_0$ has grown to $P_0 \propto \Pi_2$. According to (50) the maximum electron relativistic $\gamma$-factor at this point is

$$\gamma_f \propto \Pi_2 a_0^2 \frac{k_0^2}{k_p^2} = \frac{c\tau}{R} a_0^2 \frac{n_c}{n_e}. \qquad (56)$$

For "cigar"-like laser pulses with $\Pi_2 > 1$, wave breaking at the time (54) happens within the theory validity time (46). The wave breaking limits the maximum energy of accelerated electrons:

$$\gamma_f \propto a_0^2 \frac{k_0^2}{k_p^2} = a_0^2 \frac{n_c}{n_e}. \qquad (57)$$

The scalings (56)-(57) being specified for the bubble regime, $k_p R \approx \sqrt{a_0}$, directly lead to the energy scaling (1).

### B. Scalings for the energy conversion and the number of trapped electrons

The developed similarity theory allows us to calculate the energy accumulated in the trapped electrons $W_{tr}(t)$. The initial laser pulse energy is

$$W_{laser} = \omega_0^2 A_0^2 R^2 \tau / 4c. \qquad (58)$$

The energy of trapped electrons in the energy range between $m_e c^2 \gamma_1$ and $m_e c^2 \gamma_2$ is

$$W_{el} = m_e c^2 \int_{\gamma_1}^{\gamma_2} \gamma \, dN \qquad (59)$$

We choose the integration limits close to the maximum $\gamma$-factor of the electrons, so that $\gamma_2 = \gamma_m$ and $\gamma_1 = (1 - \beta)\gamma_m$, where $0 < \beta \leq 1$. The value of $\beta$ is a question of convention and describes a spectral width important for a particular application.

Using the expression (50) we conclude that the energy transformation efficiency is

$$\frac{W_{el}}{W_{laser}} = -P_0(t)\theta\left(\frac{k_p^2 ct}{a_0 k_0}, \Pi_1, \Pi_2, \beta\right), \qquad (60)$$

where the universal function $\theta\left(k_p^2 ct/a_0 k_0, \Pi_1, \Pi_2, \beta\right)$ depends only on the dimesionless parameters $\Pi_1, \Pi_2, \beta$ and $k_p^2 ct/a_0 k_0$.

Because of its physical meaning the universal function $\theta$ is bounded and has an upper limit

$$\overline{\lim}_{t\to+\infty}\theta\left(\frac{k_p^2 ct}{a_0 k_0}, \Pi_1, \Pi_2, \beta\right) = \Theta\left(\Pi_1, \Pi_2, \beta\right)$$

Thus, for large enough times, one can write

$$W_{tr} = -P_0(t)\Theta(\Pi_1, \Pi_2, \beta)W_{laser} \qquad (61)$$

for $k_p^2 ct \gg a_0 k_0$. Note that because of its physical meaning the function $\theta\left(k_p^2 ct/a_0 k_0, \Pi_1, \Pi_2, \beta\right)$ is monotonous and limited at large $t$. Since the largest value of $|P_0|$ is either $\Pi_2$ or 1 when the relativistic factor of the electrons is given by either Eq.(56) or Eq. (57). As a result, the energy finally transfered to the trapped electrons before the wave-breaking equals

$$W_{el} = \eta(\Pi_1, \Pi_2, \beta)W_{laser}, \qquad (62)$$

Since the transformation coefficient $\eta$ depends only on the two dimensionless parameters $\Pi_1, \Pi_2$ and $\beta$, it can be made of the order of unity by a reasonable adjustment of the laser pulse. This is also suggested by simulations [9] and recent experiments [2–4].

We point out that the Bubble regime corresponding to $k_p R \approx \sqrt{a_0}$ is parametrically different from the "blowout" regime [15], where the laser radius must be $R \leq \lambda_p$, where $\lambda_p = 2\pi/k_p$. For the blowout regime one has $\Pi_1 \leq 1/\sqrt{a_0}$, i.e. one important dimensionless parameter depends on $a_0$. According to Eqs. (62), (50), (51) and (53) the results concerning the blowout regime are not scalable to the area of large $a_0$, since the transformation energy coefficient, characteristic length, etc., have an additional nontrivial dependence on $a_0$ owing to $\Pi_1$.

If we restrict our consideration to a regime before the wave-breaking ($|P_0| < 1$) and to *tilde*-times satisfying Eq. (46), then we can evaluate the number of the trapped electrons as

$$N_{tr} = \frac{W_{el}}{\gamma_f} = \frac{a_0^{3/2}}{k_0 r_e}\left(\frac{n_c}{n_e}\right)^{1/2} D_3(\Pi_1, \Pi_2), \qquad (63)$$

where $r_e = e^2/m_e c^2$ is the electron radius. Note, that the scaling (63) can be also written as $N_{tr} \propto n_e R^3$.

For the bubble regime, $k_p R \approx \sqrt{a_0}$, the scaling (63) leads to the scaling (2).

### C. Scalings for the electron density compression

One of important practical questions for the electron acceleration is the density of the electron bunch. Because our similarity theory considers only the plasma area where electons move with ultra-relativistic energies, electron density here scales differently from the cold background. The correct scaling for the electron density compression can be extracted directly from the normalization procedure (41). Because all the *tilde*-variables are of the order of unity, one concludes that the compressed density $n_{compr}$ of ultra-relativistic electrons (those compressed by the laser pulse and those in the accelerated electron bunch) is of the order of

$$n_{compr} \propto \sqrt{a_0 n_e n_{cr}} = n_e S^{-1/2}. \qquad (64)$$

When we compare the scaling (64) with the scaling for the number of trapped electrons (63), it looks as if we had some contradiction. Indeed, at the first sight, the scaling (63) looks as if the density of the trapped electron bunch scales simply as the background density $n_e$ provided that the bunch dimensions scale together with the radius $R$ of the accelerating structure. At the same time, the formula (64) suggests that the electron density compression is $S^{-1/2}$ times stronger. To overcome this contradiction, one has to suppose that the electron bunch is *structured*. Although the length of the bunch and the electron radial excursion scale together with $R$, the bunch must have a finer internal structure on the scale of the laser wavelength $\lambda$. We abstain from a detailed discussion of this fine structure in the present paper. Yet, we mention that it is this fine structuring that is responsible for the very sharp (on the order of $\lambda$) leading edge of the accelerated bunch as observed in [9]. A fine strucutre is also seen in our 3D PIC simulation as presented in Fig. 3.

### VIII. GEOMETRY OF THE SIMILARITY

Now we want to take a closer look at the geometrical aspect of the similarity. We have formulated the similarity for the electron distribution function in the variables $\tilde{T}$, $\tilde{x}_0$ and $\tilde{\mathbf{x}}$. Thus, the plasma density distrubution describing the accelerating structure and the bunch of trapped electrons, is similar only if it is expressed in these *tilde*-variables. This geometrical similarity may be not valid in the laboratory reference frame variables $t$, $x$, $y$ and $z$. Particularly, since

$$\tilde{T} = \omega_0(t - x/c), \quad \tilde{x}_0 = Sk_0 x, \quad \tilde{\mathbf{x}} = S^{1/2}k_0\mathbf{x},$$

one easily sees that for a given $t$, different $x$ correspond to different $\tilde{T}$. In other words, consideration for a given $\tilde{T}$ means that we compare dynamics at different $x$ at different laboratoty times $t$. Generally, there is no "simultaneous" geometrical similarity as observed in the laboratory frame. Rather, the similarity is "retarded". Moreover, in the laboratory variables the coordinate $x$ has two quite different scales that are $1/k_0$ and $1/Sk_0$ respectively. As a result, the picture of the motion in these variables looks much more sophisticated.

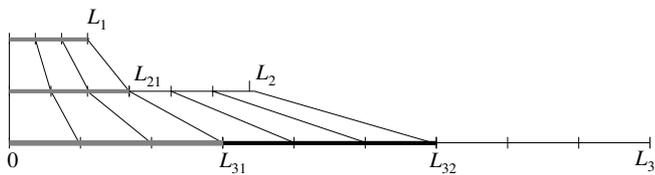

FIG. 2: A schematic illustration of the ladder similarity (see text).

The last point we are going to discuss in detail is the non-trivial structure of the developed similarity theory. There are two characteristic lengths in the laser propagation direction. One scale is $L^{sim} \propto a_0 k_0/k_p^2$. It is defined by the normalization (41). Another scale is given either by the wave breaking distance $L_b \propto L^{sim} S^{-1/2}$ (53) or by the laser depletion length $L_d = \Pi_2 L_b$ for $\Pi_2 \leq 1$. Our similarity theory states that if we have two laser-plasma interaction cases with $S_2 < S_1$, then the second interaction dynamics is similar to the first one on the distance $L_2^{sim} = (S_1/S_2)L_1^{sim}$. Yet, the laser pulse in the second interaction runs for a longer distance $L_2 \propto L_2^{sim} S_2^{-1/2}$. This new interval in the second interaction cannot be scaled up from the first interaction with a smaller $S$-parameter. Thus, the interactions contain similar and non-similar parts, and the similarity is "ladder"-like.

The ladder similarity is schematically demonstrated in Fig. 2. Let us have three cases of laser-plasma interaction with $S_1 > S_2 > S_3$. Then, the interaction 2 is similar to the interaction 1 up to the distance $L_{21} = (S_1/S_2)L_1$. The interaction between the distances $L_{21}$ and $L_2 \propto (S_1/S_2)^{3/2} L_1$ is new. In the same manner, the interaction 3 is similar to the interaction 1 up to the distance $L_{31} = (S_1/S_3)L_1$. Further, the interaction 3 is similar to the interaction 2 up to the distance $L_{32} = (S_2/S_3)L_2$, and so on.

## IX. NUMERICAL CHECK OF THE ANALYTICAL SCALINGS

To check the analytical scalings, we use 3D Particle-in-Cell simulations with the code VLPL (Virtual Laser-Plasma Laboratory) [10]. In the simulations, we use a circularly polarized laser pulse with the envelope $a(t, \mathbf{r}_\perp) = a_0 \cos(\pi t/2\tau) \exp(-\mathbf{r}_\perp^2/R^2)$, which is incident on a plasma with uniform density $n_e$. Here, $a_0 = eA_0/mc^2$ is the dimensionless laser amplitude.

First, we check the basic ultra-relativistic similarity with $S = const$. We choose the laser pulse duration $\tau = 8 \cdot 2\pi/\omega_0$, where $\omega_0$ is the carrier frequency of the laser pulse. The laser radius is $R = 8\lambda$, where $\lambda = 2\pi c/\omega_0$ is the laser wavelength. The parameter $\Pi_2 = c\tau/R = 1$ in this case.

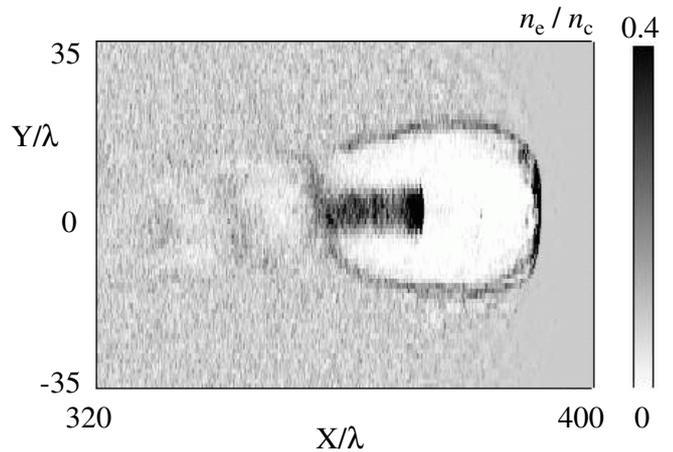

FIG. 3: Electron density taken from the 3D PIC simulation (iv) with parameters $a_0^{iv} = 80$, $n_e^{iv} = 0.08$ after the laser has propagated the distance $L = 400\lambda$ in plasma. The wake field takes the form of a single cavity (the Bubble) [9], free from cold background electrons. The dense bunch of trapped and accelerated electrons is structured and has a sharp leading edge.

We fix the basic similarity parameter to the value $S^i = 10^{-3}$ and perform a series of four simulations with (i) $a_0^i = 10$, $n_e^i = 0.01n_c$; (ii) $a_0^{ii} = 20$, $n_e^{ii} = 0.02n_c$; (iii) $a_0^{iii} = 40$, $n_e^{iii} = 0.04n_c$; (iv) $a_0^{iv} = 80$, $n_e^{iv} = 0.08n_c$. Assuming the laser wavelength $\lambda = 800$ nm, one can calculate the laser pulse energies in these four cases: $W^i = 6$ J; $W^{ii} = 24$ J; $W^{iii} = 96$ J; $W^{iv} = 384$ J. These simulation parameters correspond to the bubble regime of electron acceleration [9], because the laser pulse duration $\tau$ is shorter than the relativistic plasma period $\sqrt{a_0}\omega_p^{-1}$. Electron density distribution obtained in the simulation (iv) after the laser pulse has propagated the distance $L = 400$ $\lambda$ is shown in Fig. 3. The wake field takes the form of a solitary cavity (the bubble) free from the cold electrons. The bubble runs with the group velocity of the laser pulse. It traps electrons from the background and accelerates them to high energies. We let the laser pulses propagate the distance $L_b^i = 1000$ $\lambda$ through plasma in the all four cases. At this distance, the laser pulses are depleted, the acceleration ceases and the wave breaks.

Fig. 4(i)-(iv) shows evolution of electron energy spectra for these four cases. One sees that the energy spectra evolve quite similarly. Several common features can be identified. First, a monoenergetic peak appears after the acceleration distance $L \approx 200$ $\lambda$. Later, after the propagation distance $L \approx 600$ $\lambda$, the single monoenergetic peak splits into two peaks. One peak continues the acceleration towards higher energies, while another peak decelerates and finally disappears. Comparing the axises scales in Fig. 4, we conclude that the scalings (14) and

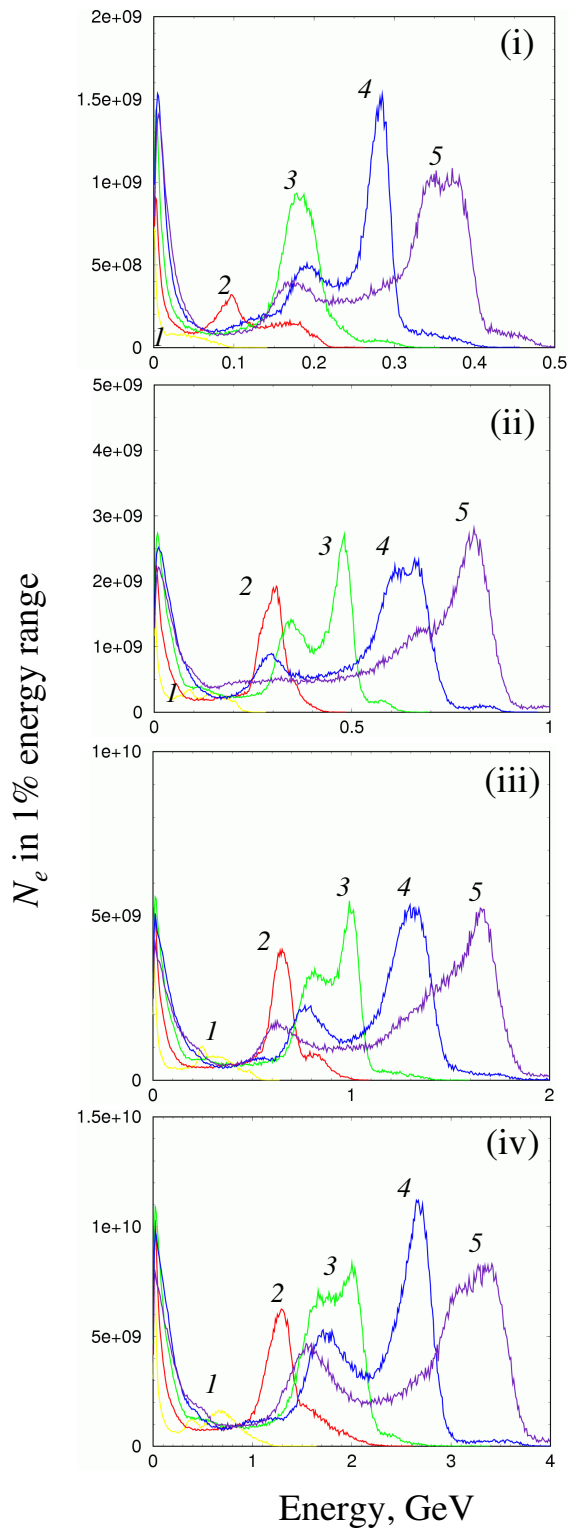

FIG. 4: Electron energy spectra obtained in the simulations (i)-(iv) (see text). The control points $1-5$ were taken after the propagation distances $L_1 = 200\lambda$, $L_2 = 400\lambda$, $L_3 = 600\lambda$, $L_4 = 800\lambda$, $L_5 = 1000\lambda$. The spectra evolve similarly. The monoenergetic peak positions scale $\propto a_0$ and the number of electrons in a 1% energy range also scales $\propto a_0$ in agreement with the analytic scalings (14)-(15).

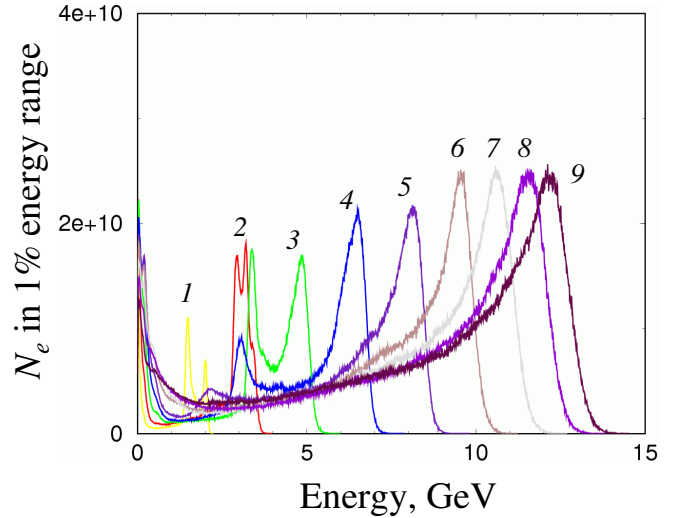

FIG. 5: Electron energy spectra obtained in the simulations (v) (see text). The control points $1-9$ were taken after the propagation distances $L_1 = 800\lambda$, $L_2 = 1600\lambda$, $L_3 = 2400\lambda$, $L_4 = 3200\lambda$, $L_5 = 4000\lambda$, $L_6 = 4800\lambda$, $L_7 = 5600\lambda$, $L_8 = 6400\lambda$, $L_9 = 7200\lambda$. The spectral evolution for the control points $1-5$ is similar to that of the simulation cases (i)-(iv). The spectra $6-9$ correspond to a new evolution that cannot be directly scaled from the previous simulations.

(15) hold with a good accuracy.

Now we are going to check the similarity theory for the slow dynamics. For this purpose, we change the $S-$parameter. We choose the laser amplitude $a_0^{\mathrm{v}} = 80$ and the plasma density $n_e^{\mathrm{v}} = 0.02 n_c$. This corresponds to $S^{\mathrm{v}} = 2.5 \cdot 10^{-4}$ and the laser energy $W^{\mathrm{v}} \approx 1.5$ kJ;. In this case, the initial laser radius and duration must be increased by the factor $\sqrt{S^{\mathrm{i}}/S^{\mathrm{v}}} = 2$. Thus, we use the laser pulse with $R^{\mathrm{v}} = 16\lambda$ and $\tau^{\mathrm{v}} = 16 \cdot 2\pi/\omega_0$. This case gives the pure density scaling when compared with the case (iv), or the pure laser amplitude scaling when compared with the case (ii). We let the laser run $L_b^{\mathrm{v}} = 8000\lambda$ through the plasma. At this distance, the energy of the laser pulse is completely depleted and the wave breaks. The change of the wavebreaking length $L_b^{\mathrm{v}}/L_b^{\mathrm{i}} = \left(S^{\mathrm{i}}/S^{\mathrm{v}}\right)^{3/2}$ coincides with the scaling (53).

The electron spectrum evolution obtained in this simulation is shown in Fig. 5. The spectra display a monoenergetic peak whose energy continuously grows up to some 12 GeV at the end. Between the control points, where the spectra in Fig. 5 have been taken, the laser pulse propagated the distance $L = 800\lambda$. This distance between the control points is $S^{\mathrm{i}}/S^{\mathrm{v}} = 4$ times larger than that in the cases (i)-(iv). One sees that the first five electron spectra in Fig. 5 are similar to those in Fig. 4. However, the last four spectra in Fig. 5 are new. This corresponds to the ladder similarity as discussed in section VIII.

## X. CONCLUSIONS

In this section we recapitulate the most important results of the developed similarity theory for ultra-relativistic laser plasmas and provide some simple "engineering" scalings for electron acceleration in the Bubble regime.

(i) Ultra-relativistic laser-plasma interactions with different $a_0$ and $n_e/n_c$ are similar as soon as the similarity parameter $S = n_e/a_0 n_c = const$ for these interactions. In this case, electron energies scale as $E_e \propto a_0$ and the number of accelerated electrons also scale as $N_{tr} \propto a_0$.

(ii) The electron dynamics in relativistically underdense plasmas, $S \ll 1$, can be separated into a fast dynamics and a slow one. The fast dynamics corresponds to a relaxation to some attractor solution. Its characteristic time is $\Delta t_f \propto k_0/(ck_p^2)$. The slow dynamics describes an adiabatic evolution of this attractor on the characteristic time scale $\Delta t_s \propto a_0 k_0/(ck_p^2)$.

(iii) The slow electron dynamics leads to a dimensionality reduction and to a remarkable 3D wave breaking exclusion rule (35). Thus, the wave breaking in the realistic 3D geometry is qualitatively different from that in the 1D case.

(iv) Starting with the Maxwell equations on the fields and the relativistic Vlasov equation on the electron distribution function we were able to derive scalings for the electron acceleration. Here, we list the most important scalings once again. The similarity theory tells us that the "equilibrium" radius $R$ of the laser pulse scales as

$$k_0 R \propto S^{-1/2}.$$

For the acceleration to be efficient and to generate structured (quasi-monoenergetic) electron energy spectra, the pulse duration $\tau$ must be

$$\tau \leq R/c.$$

The characteristic acceleration length $L_b$, i.e., the length before the wave breaking happens, scales as

$$k_0 L_b \propto \frac{c\tau}{R} S^{-3/2}.$$

The maximum electron energy $E_{acc}^{max}$ achieved in this interaction scales as

$$E_{acc}^{max} \propto \frac{c\tau}{R} a_0 S^{-1}.$$

The number of electrons accelerated to this energy scales as

$$N_{tr} \propto \frac{1}{k_0 r_e} a_0 S^{-1/2}.$$

Using the analytical theory and the numerical simulation results we can express the maximum electron energy $E_{acc}^{max}$ (measured in GeV) as a function of the laser pulse energy $W_{laser}$ (measured in Joules). First, we consider the case when we fix the similarity parameter to some constant, e.g., $S = 10^{-3}$. In this case, the energy scaling is

$$E_{acc}^{max}[\text{GeV}] \approx 0.17 \left(W_{laser}[\text{J}]\right)^{1/2}.$$

Thus, the electron energy scales as a square root of the laser energy. One increases simultaneously the laser pulse amplitude and the plasma density, while the acceleration length remains constant $L_{acc} \approx 1$ mm. A faster scaling for the electron energy can be obtained, if one fixes the laser amplitude, e.g., $a_0 = 10$, fixes the pulse aspect ratio $c\tau/R \approx 1$ and varies the plasma density $n_e$. In this case, the energy scaling is

$$E_{acc}^{max}[\text{GeV}] \approx 0.1 \left(W_{laser}[\text{J}]\right)^{2/3}. \quad (65)$$

At the same time, the acceleration distance grows as

$$L_{acc} \approx 0.03 \cdot S^{-3/2} \, \lambda,$$

where the factor 0.03 is taken from the simulation.

The scaling (65) means that to build an efficient laser-plasma electron accelerator in the 100 GeV energy range, one would need a 10 kJ class laser with the pulse duration of some 100 fs.

In the end, in this work the very first analytical results for ultra-relativistic 3D laser-plasma interaction are obtained. The developed non-trivial similarity theory will help to make another step towards a practical realization of a high-energy laser-plasma electron accelerator.

### Acknowledgments

This work was supported in parts by the Transregio project TR-18 of DFG (Germany) and by RFFI 04-02-16972, NSH-2045.2003.2 (Russia).

---